\documentclass{article}

\usepackage[final]{nips_2016}

\usepackage[utf8]{inputenc} 
\usepackage[T1]{fontenc}    
\usepackage{hyperref}       
\usepackage{url}            
\usepackage{booktabs}       
\usepackage{amsfonts}       
\usepackage{nicefrac}       
\usepackage{microtype}      

\title{ABM: an automatic supervised feature \\engineering method for loss based models \\based on group and fused lasso}

\begin{document}

\maketitle

\section{Author}
\textbf{Weijian Luo}\\
Department of Mathematcial Science\\
Peking University\\
Beijing, 5th Yiheyuan Road\\
\texttt{1701210055@pku.edu.cn}

\textbf{Yongxian Long}\\
Wisecotech Corporation\\
Beijing\\
\texttt{longyongxian@eisecotech.com} \\

\begin{abstract}
	A vital problem in solving classification or regression problem is to apply feature engineering and variable selection on data before fed into models.One of a most popular feature engineering method is to discretisize continous variable with some cutting points,which is refered to as bining processing.Good cutting points are important for improving model's ability, because wonderful bining may ignore some noisy variance in continous variable range and keep useful leveled information with good ordered encodings.However, to our best knowledge a majority of cutting point selection is done via researchers domain knownledge or some naive methods like equal-width cutting or equal-frequency cutting.In this paper we propose an end-to-end supervised cutting point selection method based on group and fused lasso along with the automatically variable selection effect.We name our method \textbf{ABM}(automatic bining machine). We firstly cut each variable range into fine grid bins and train model with our group and group fused lasso regularization on each successive bins.It is a method that integrates feature engineering,variable selection and model training simultanously.And one more inspiring thing is that the method is flexible such that it can be taken into a bunch of loss function based model including deep neural networks.We have also implemented the method in R and open the source code to other researchers.A Python version will also meet the community in days. 
\end{abstract}

\section{Inroduction}

\subsection{Feature Egineering and Bining}
A lot of machine learning models have been proposed to solve classification and regression problems in past decades.Some models gain great success in certain fields.However, directly use of statistical models could often lead to catastrophies like poor accuaracy or less interprataition ability. A more elegent solution to fix the problem is to do some feature engineering and variable selection before getting final frozen model.A typical case reveals importance of feature engineering happens when we consider data with both continous and discrete variable as features.Most model fitting dynamics only focus on one type of features, continous of discrete.So we have to either discretisize the continous on to serveral levels or we continousize the discrete on to some high dimenstional space like one-hot encoding does. In this paper we focus mainly on the continous cases.

How to discretisize a continous variable into some level is of important status in feature engineering fields.For simplicity, we name the discretized variable "bined variable",and its corresponding discrete levels as "bins".A good bining method will significantly improve model fitting ability in two aspects.One aspect is that good bins may ignore noisy variation for continous variables which bring higher signal to noise ratio.Another aspect is that good bining will reduce values taken for one variable, which can avoid value sparsity problem which may heart training process and demanding "more" data. A lot of success feature engineering method has been proposed in past years. Case like HOG feature in computer vision fields have dominanted the face recognition technology before deep learning era.The kernel trick for support vector machine can also been seen as a feature engineering case because the kernel framework implicitly mapping the original features into some invisible feature space where information from features could be expressed with more power.There are also some cases in big data credit scoring fields.Empirical experiments have show that woe(weight of evidence) encoding for a scorecard model training can enhance model accuracy.  

Despite the importance of feature bining, most bining methods in different fields relies heavily on researcher's domain knowledge.Such examples happens a lot such like in variable "age", a typical bining stategy is to combine age from 1 to 7 as one bin, 7 to 18 as second one and so on.But easy to see, the bining stategy is both not optimal and unsupervised so in some modelling cases this bining strategy may lead to decrease of model accuracy.Other popular feature bining methods are equal-width and equal-frequency bining method which choose cutting splits to cut the variable range into some intervels with equal length.While the equal-frequency bining method choose bining splits according to variable quantiles such like Q1 to Q10.Although these two bining method is widely used, they are both unsupervised and with very naive motivation.These method may lead to bad bins in some cases such like the case when variable taking values near some finite centors.However, naive method does not bring nothing.In later work we use fine grid equal-frequency bins as an initialization for auto-bining algorithm.Another popular method for automatic bining is the decision tree based methdo.\textbf{WHO} propose to use decision tree as a bining alogorithm and run tree training to obtain final cut points for variables.Though this method have gained some sucess, it is critisized for its overfitting and lack of explaination.Another disadvantage of decision tree based method is that the tree model gives sequential cut operation between variables which means one variable bin is highly related to its predecessor variable and the sequence order may be determined randomly.The sequential structure for model bining may lead to some instability in some problem which we do not prefer.There are yet another some bining method like chisquare bining method which is based on high dimensional contengence table test.These methods often do not give visible improvement for model accuracy and generalization ability so they are not used widely.

\subsection{Loss Based Training and Regularization}
As we mentioned before there are huge bunch of models proposed from different background,statistics or geometry,a typical paradigm goes along with them.These models' training can all be formulated as a mathematical optimization problem.The objective of the problem is often refered to as \textbf{loss function}.For example,the Linear regression model training targets on optimizing the loss function $loss = \Sigma ||x_i^T \beta - y_i||_2^2$,logistic regression model optimize function $loss = \Sigma [y_i \beta^T x_i - log(1+e^{\beta^T x_i})]$,even the most popular deep neural network model applied in many fields could be abstracted to a sophiscated model architecture and a specially designed loss function to train model.By optimizing the loss function,we get the estimation of model parameter.And with model parameter, we could inference new input sample to its response.

Based on loss function paradigm, regularization term is often added to the loss function to apply joint optimization on estimating parameters.The ridge regression loss was proposed early in 1980s, which could be viewed as a l2 norm regularized linear regression loss. Tibshirani proposed the \textbf{LASSO} estimator for linear regression model parameters,which could be viewed as a l1 norm regularization on linear regression loss. \textbf{WHO} proposed the elastic net regression loss, which could be viewed as a combination of both l1 and l2 norm regularization.The introduction of regularization term is of good benefit on many aspects.Some regulariztion terms could be interpreted as a bayes statistics prior distribution factor to likelihood, and algorithmically the regularization term will help the optimization problem lead to a shrinkage or sparse estimation,which may bring more insight over data.

Another background we are delivering is the bin logistic regression model. The classical 2 class logistic regression model takes a p dimensional input feature $\vec{x} = (x_1,...,x_p)$, and output a probability that the input $\vec{x}$ belong to the positive class with $\frac{1}{1+e^{\beta^T\vec{x}}}$.The logistic regression model directly use original variable values to train the model.Based on logistic regression, a updated version of logitsic regression model is the bin logistic regression model.Unlike the logistic regression, bined logistic regression model divide one continous variable range into several  ordered finite or infinite intervals with no intersection, which is refered to as bins.And then the model encoded each bins to a one-hot like encoded 0-1 sub-variable and accecpt these bined sub-variables as input. The charming part of bined logistic regression is that bins could carry risk trends information and thus the bined model could be explainable in a human way.

Of course, the most challenging problem in bined statistical model is the bining methods wihch tell us how to discretize variables' range into some ordered bins.Equal-width or equal-frequency bining methods are traditionally used but these two methods are too naive or a complicated model fitting. In recent years, \textbf{WHO} has proposed a chisquare method for variable bining, and \textbf{WHO} have proposed a decision tree based method for variable bining.The chisquare bining methods is supposed to have asymptotic convergence property but does not work well in cases when sample size is not that large.The decision tree based bining methods use decision tree dynamic to do variable bining but it bins variables sequentially one by one and may lead to an greedy local optimal bining result which is not strongly recommanded.In industry fields, many models are hand-crafted bined using reseacher's field knowledge and inpiration.

Bin logistic regression model have gained great applications in credit risk scoring fields with a special name,the scorecard model. Scorecard model have been very popular in risk scoring scenes due to its simplicity, flexibility on dealing with both continous and discrete variables and, most importantly, its explainability.

In this paper, we propose a group and group fused lasso regularization for bin logistic regression model training. The method will achieve a simple and end-to-end way for loss based models to automatically find optimal cutting point to discretize input variables into finite bins, along with the effect of variable selection. The automatically bining method is of great novelty because it avoid the hand-crafted bining process which is non-data-driven or less data-driven.On the contrary, the automatically bining algorithm is strongly data-driven and is supervised which may lead the model to learn more insight about data.What is more is that our method integrates the global information about input variables simultanously, while the hand-crafted method only watch the marginal information for one variable at one time and the decision tree based bining algorithms consider a sequential variable bining which may lead to a local optimal bining instead of a global optimal bining.

\section{Related Work}
\subsection{Binnig and Scorecard Model}
Scorecard model is a certain type of statistical model for classificatio problems and is widely used in big data risk scoring field.Scorecard model is very similar to bined logistic regression and can be even treated as a special case of bin logistic regression. The little difference lies in a small change of loss function between scorecard and bin logistic regression model.The bin logistic regression model usually use a negative log bernoulli likelihood as loss function while the scorecard model always use divergency as a loss function to optimize.To train a scorecard model, you are supposed to bin continous variable to few intervals,named bins. You may apply hand-crafted marginally on each variable separately or use some algorithm to automatically bin. After the bining process, supposing you have made n bins, each of your continous feature could be located in certain one bin, and thus be one-hot encoded to n binary sub-variables with each one representing whether the feature $x_{i,j}$ locates in its bin. Then the scorecard model eat these sub-variables as input and train a logistic regression model with some loss functions. Loss function could be cross-entropy or divergency.Our implemention of automatical bining algorithm later use scorecard model as base model over some classical open source credit risk modelling dataset.We choose scorecard model as base model because of its popularity and domain adaptiveness.Our method can be implemented over huge bunch of loss based models with great flexibility.

\subsection{Fused Lasso}
\textbf{Fused Lasso} was proposed by \textbf{Tibshirani} at first time.Fused Lasso is a regularization term with which penalize sequentially variables.Suppose you have one sample $(\vec{x_i},y_i)$ and $\vec{x_i} = (x_{i,1},...,x_{i,p})$.A fused lasso regularization term is formulated as $$flasso = \lambda \Sigma_{2 \leq j \leq p} |\beta_j - \beta_{j-1}|$$

The fused lasso regularization term is a penalization for variables  in succession.This penalization will force the paramter of two successive variable $\beta_{j-1}$ and $\beta_j$ to be exactly the same or have a visible gap.And a wanderful insight about fused lasso regularization is that the regularization term will force the successive two variables' parameter to have same level or have a jump between them.And hopefully it may happen that the successive k variables have exact the same coefficient. In an open view, the phenomenon that successive k variables having same coeficient could be seen as that these variables contribute same effect to response and may be combined in some way to reduce the model complexity.This combination thought will give rise to an automatic bining dynamic which we will talk later.

Wide application of fused lasso has also been proposed frequently. \textbf{WHO} proposed to use fused lasso method to detect change point of a sequence data or time series. \textbf{WHO} proposed to generalize the fused lasso method to graph fused lasso method to capture graphical correlation between variables.A lot of algorithmic imrpovements have also been proposed.\textbf{WHO} gives an optimization alogorithms to use a primal-dual paradig optimization method to efficiently solve fused lasso problems.And \textbf{WHO} have proposed a use of Alternating direction of multiplier method to efficiently solve the fused lasso problem. 
\subsection{Group Lasso}
\textbf{Group Lasso} was proposed by \textbf{WHO} at first time.Group lasso is a regularization term which could influence the regression coefficient estimation to simultanously keep or droup certain group of variables.More precisly, assume we have k groups of variables $\vec{x_1},..,\vec{x_k}$.Each $\vec{x_i}= (x_{i,1},...,x_{i,p_i})$ represents a group of variable.The group formulation is often very natural in specific problem.For instance, one man's features could be coarsely classified to some typical module such as profiles, behaviours and other modules.So one man's feature variables could be naturally splitted to serveral groups of variables.Each group variables represent a typical aspect of one man's feature.Armed with a good split of groups, the group lasso regularization term could be formulated as 

$$glasso = \Sigma_i \lambda_i ||\vec{\beta_i}||_2$$
Where $\vec{\beta_i}$ is the ith \textbf{group} coeficients vector corresponding to the ith \textbf{group} of features.Easy to see, the group lasso term is just an sum of l2 norm of each grouped features.This simple regularization term penalize \textbf{group} of variables.The penalization could force one group of variables to exists or simultanously have zero coefficients.This could be interpreted as an automatic way of group preserveing or group dropping of variables.This dynamic could realize a \textbf{group} variables selection effect and thus being wildly studied.\textbf{WHO} proposed to apply the grouped lasso on medical diagnosis problem and acheive a rubust model.\textbf{WHO} proposed to use an L1 group penalty instead of an L2 group penalty. Some theoretical analysis have also been proposed successively.\textbf{WHO} studied some properties for group lasso methods.\textbf{WHO} give an statistical explaination of group lasso.Numerical optimization algorithms are also proposed actively.\textbf{WHO} propose. 
\subsection{Group and fused lasso}
The idea of introduce both group lasso and fused lasso into logistic regression training has not been studied widely.\textbf{WHO} proposed to use group and fused lasso to \textbf{Do Something}.\textbf{WHO} give a theoretical analysis for group and fused lasso method.

\section{Our Proposal}
In this paper, we deliver an end-to-end methodololy which could output an explainable data-driven model for classification problems. Our method is based on bin logistic regression model or scorecard model, and we equiped the trianing loss for model with a group lasso term and a group fused lasso term. That is we propose to train an bin logistic regression model with a group lasso and a fused group lasso regularized logistic loss function.The proposal method is very flexible for a lot of model training, not constrained to bined logistic regression model. Models like orther additive models could also gain performance elevation via this training method. The proposal achieves the effect of automatical bining with no domain knowledge while training the model. The detailed formulation of methods is following. Suppose we have $(x_i,y_i), 1\leq n$ as dataset, where each feature vector $\vec{x_i}$ consists of $p$ variables realization, which means $$\vec{x_i} = (x_{i,1},...,x_{i,p})=(x_{i,j}),1\leq j \leq p$$.

We do not directly use \textbf{raw} variable as input because of two reasons.The first reason is that the mapping from one continous value variable to target may be hard to be learn because of its sparsity in values taken.The second reason is that the interpretation for contribution of a raw continous variable to prediction is hard to be shown. Instead we apply a coarse bins operation at first.What is a coarse bin? For one variable $x_{i,j},1\leq i \leq n$, we divide the variable range to $nbins$ ordered intervals.For instance, if one variable $x_{i,j},1\leq i \leq n$
takes values in range from 0 to 100, then we could divide the range $[0,100)$ into $nbins=10$ intervals with equal width, and the bin intervals is supposed to look like $[0,10),[10,20),...,[90,100)$. We use notation $\mathcal{B}$ to denote the bins set $\mathcal{B} = \{B_1,..,B_{nbins} \}$ , where $B_i$ is the exact ith bin interval. Having bin intervals for one variable, we encode raw continous variable to an $nbins$ dimensional binary embedding feature with mapping $x_{i,j}^{emb} = (\mathbb{I}_{\{x_{i,j} \in B_1\}},...,\mathbb{I}_{\{x_{i,j} \in B_{nbins}\}})$.It is clear to see we extend the original variable to a $nbins$ dimensional feature vector consists of sub-variables. We encodes $p$ variables into $p\times nbins$ sub-variables. Each sub-variables represents a \textbf{level} of effect the variable could contribute to the target.

After we coarse bin the input variables to sub-variables, one may choose to combine neighbour bins as larger bins with domain knowledge and tuning techniques.But our method could achieve automatic combination of bins and automatic variable selection simultanously.We formulate the automatic bining and automatic variable selection problems as a group lasso and group fused lasso problem.Suppose we have $p$ groups of sub-variables encoded from $p$ original variables.Sub-variable is denoted by 
$$x^{sub}_{i,j,k},1\leq i \leq n,1\leq j\leq p,1\leq k \leq nbins$$
where i represents the i the observation, j represents the jth group of sub-variables and k represents the kth sub-variable in jth group.We denote the target as $y_i$ which is a binary response taking value from $\{0,1\}$. It is supposed that the conditional distribution of $y_i | x^{sub}_i$ follows a bernoulli distribution 
$$ P(y_i=1|x^{sub}_i) = \frac{1}{1+e^{\beta_0 + \Sigma_{i,j,k} \beta_{i,j,k} x^{sub}_{i,j,k}}}$$
$$ P(y_i=0|x^{sub}_i) = \frac{e^{\beta_0 + \Sigma_{i,j,k} \beta_{i,j,k} x^{sub}_{i,j,k}}}{1+e^{\beta_0 + \Sigma_{i,j,k} \beta_{i,j,k} x^{sub}_{i,j,k}}}$$
and it is supposed further joint variables $(x_i,y_i),1\leq i \leq n$ are independent.The model parameter $\beta_{i,j,k}$ and $\beta_0$ is the coeficient of each bin.

The likelihood function of $y|x$ is
$$P(y|x) = \Pi_i (\frac{1}{1+e^{\beta_0 + \Sigma_{i,j,k} \beta_{j,k} x^{sub}_{i,j,k}}})^{y_i}(\frac{e^{\beta_0 + \Sigma_{i,j,k} \beta_{j,k} x^{sub}_{i,j,k}}}{1+e^{\beta_0 + \Sigma_{i,j,k} \beta_{j,k} x^{sub}_{i,j,k}}})^{1-y_i}$$
While the negative log-likelihood function 
$$ Loss = -log(P(y|x))$$ 
is often viewed as a loss function to train a logistic regression model. Which means the logistic regression parameter $\beta_{i,j,k}$ is estimated by

$$\hat{\beta}_0,\hat{\beta}_{i,j,k} = argmin_\beta -log(P(y|x))$$

We formulate the automatic bining problem as a constrained optimization problem based on logistic regression fitting.That is we propose a optimization problem:

$$\hat{\beta}_0,\hat{\beta}_{i,j,k} = argmin_\beta -log(P(y|x))/n + \lambda_1 \Sigma_j \Sigma_{1\textless k \leq nbins} |\beta_{j,k+1}-\beta_{j,k}| + \lambda_2 \Sigma_j \Sigma_k \sqrt{x_{j,k}^2} $$
It can be simplified as 
$$\hat{\beta}_0,\hat{\beta}_{i,j,k} = argmin_\beta -log(P(y|x))/n + \lambda_1 \Sigma_j ||diff(\vec{\beta_j})||_{1} + \lambda_2 \Sigma_j ||\vec{\beta_j}||_2 $$
where the $\vec{\beta_j}$ is the jth group of parameters and $\lambda_1$,$\lambda_2$ are two tuning parameters.

The novalty inside this formulation is that we can use tuning parameters $\lambda_1$ and $\lambda_2$ to control strength of automatic bining and variable selection effect respectively.The regularization term $\Sigma_j \Sigma_{1\textless k \leq nbins} |\beta_{j,k+1}-\beta_{j,k}|$ force the successive two bins' coefficient to be exactly the same or have a visible gap, which could naturally be interpreted as a combination of bins if coefficients of two successive bins are the same.Another piece of novalty is that the regularization term $\Sigma_j ||\vec{\beta_j}||_2$ could force one group of sub-variables to have zero coefficients simultanously or not.If one group of sub-variables have exactly zero coefficients, it is naturally interpreted that the variable derived the group of sub-variables is supposed to be dropped from the model.The group lasso and group fused lasso method considers global information among all variables simultanously, especially it takes pairwise interaction information between each two pairs of variables.

The method integrates the advantage of fused lasso and group lasso with bin logistic regression.The method could also been generalized to other machine learning algorithms like support vector machine or GLMs.The method could also been treated as a trick for deep neural network models.

\section{Experiment}
We apply experiments on serveral open dataset and some simulated data.We compare our end-to-end model training pipeline with automatic bining with default logitic regression,equal-frequency bining, equal-width bining, logistic regresion with decision tree bining.The automatic bining method has shown great advantage over rest algorithms in both performance and explaination ability.

\section{Conclusion}

\section*{References}

\medskip

\small

[1] R. Tibshirani, \ \&M. Saunders, \ \&S. Rosset, \ \&J. Zhu\ \&K. Knight.\ (2005) Sparsity and smoothness via the fused lasso. {\it Journal of the Royal Statistical Society: Series B} {\bf  67}(1):91–108.

[2] R. Tibshirani, \ \&M. Saunders.\ (2009) Properties and Refinements of the fused lasso {\it The Annals of Statistics} {\bf  37}(5B):2922–2952.

[3] J. Friedman, \ \&T. Hastie, \ \&R. Tibshirani.\ (2010) A note on the group lasso and a sparse group lasso

[4] The group lasso for logistic regression

[5] On Change Point Detection using the fused lasso method

[6] The group fused Lasso for multiple change-point detection

[7] Regularized Estimation of Piecewise Constant Gaussian Graphical Models: The Group-Fused Graphical Lasso

[8] Sparse Group Fused Lasso for Model Segmentation

[9] Model selection and estimation in regression with grouped variables

[10] The Study of Credit Scoring Model Based on Group Lasso

[11] Fused Lasso for Feature Selection using Structural Information

[12] An error bound for Lasso and Group Lasso in high dimensions

[13] Correlated Feature Selection with Extended Exclusive Group Lasso

[14] Fused least absolute shrinkage and selection operator for credit scoring

\end{document}